\documentstyle[12pt]{article}
\def\ii{\'{\i}}
\def\bi{\bigskip}
\def\noi{\noindent}
\def\be{\begin{equation}}
\def\en{\end{equation}}
\def\bq{\begin{eqnarray}}
\def\eq{\end{eqnarray}}
\begin{document}

\begin{center}
{\Large \bf ON THE SIGN OF THE STATIC DIELECTRIC FUNCTION AT $T=0$}
\\[1.5cm]
\end{center}
\begin{center}
{\large\bf Alejandro Cabo Montes de Oca\footnote{Work supported 
by CONACyT under contract 3979PE-9608}}\\[.3cm]
{\it Abdus Salam International Centre for Theoretical Physics}\\
{\it Miramare, Trieste, Italy,}\\
\vspace{.5cm}
{\it Instituto de F\'\i sica, Universidad de Guanajuato}\\
{\it Lomas del Bosque \# 103, Lomas del Campestre}\\
{\it 37150 Le\'on, Guanajuato; M\'exico}\\
{\it and}\\
{\it Instituto de Cibern\'etica Matem\'atica y F\ii sica}\\
{\it Calle E, No. 309. Esq. a15, Vedado, La Habana, Cuba}
\end{center}
\noindent

\vspace{0.1cm}

\begin{center}
{\bf Abstract.} 
\end{center}

The question of allowed signs of the static dilectric function for exactly     
homogeneous ground states in many body systems is further analyzed. The  
discussion is restricted to zero temperature situation. Firstly, it is argued that
the positivity is equivalent to the requirement that the work needed to  
adiabatically connect an static charge density, be positive. In particular 
the rule seems to be a fully appropiate stability condition if the system 
interact with a freely moving but Coulomb interacting compensating jellium. 
This situation is common in thermodynamical equiliubrium discussions,
in which the system can 
be stable or not in dependence of the relaxation of some external constraints.
Further, an argument based in a quantized electromagnetic field treatment is also
given. It requires $\epsilon > 0 $ as an intrinsic stability condition for the
charged system independently of the jellium dynamics. Possible implications for the
more complex finite temperature situations are commented.  

\setlength{\baselineskip}{1\baselineskip}

\newpage

\section{Introduction.} 

\bi
\bi 

The problem of the positivity of the static dielectric function have been
debated in the literature in close connection with a related but not  
equivalent question about the validity of the dispersion relations [1-4]. In
the works [3] and [4] it has been argued that negative values of $\epsilon 
(\vec k)$ are allowed by stable translation invariant physical systems.
 The conclusion was extracted from
a discussion of appropriate effective actions which should have minimum values
at equilibrium.

\bi

In a previous work [5] we have considered the requirement for a positive work
in adiabatically creating a fluctuation of the charge density in the 
background. The conclusion was that such a condition implies the
positivity
of $\epsilon (k) $. Then, in this work we intend to clarify the 
connections among the results of [3],[4] and our work [5]. Only the null 
temperature situation will be considered here. The harder physical problematic
at $T \not = 0$ will be discussed elsewhere.

\bi

The work start in Section 1, from adapting the variational
 derivation of the positivity
of the
effective action, to the problem [6]. The considered physical system consists
of several Coulomb interacting species of charged particles in interaction  
with a compensating jellium. Then, the adiabatic connection of the auxiliary
electrostatic external source allow to obtain the same stability conditions of
[3] and [4] from the positivity of the effective action. The variational 
discussion being used  clarifies the meaning of the positivity for the 
work done in 
connecting the charge functuation [5]. It is simple equivalent to requiring
that the energy $E[\rho ]$ of the state after the adiabatic connection of the
sources becomes greater than the ground state energy $E_g$. The inequality 
$\epsilon (\vec k)\geq 0$ directly follows from this requirement.

\bi

A first conclusion then follows. It refers to systems in
which the jellium charge density is freely moving but Coulomb interacting 
with the particles, that is, systems in which the jellium homogeneity
is 
assured self consistently by the Coulomb interaction. In such situation if 
$E[\rho ]$ can be lower than the ground state energy $E_g$, the system can 
relax by itself to an inhomogeneous configuration by generating a jellium 
density modulation. Therefore, in order that the whole system: particles + 
jellium, be stable the positivity of $\epsilon (\vec k)$ is needed.

 Next, in Section 2, after considering the  electromagnetic field
as
quantized, the positivity of $\epsilon(\vec k)$ follows as an intrinsic
stability
requirement coming from the internal dynamics of 
the charged particle system and being independent of the jellium nature.   

 In the summary, the main conclusions are reviewed  and comments  on the
posible implications for the  finite temperature case are given.
\bi

\section{ $\epsilon(\vec k) >0$ and positive work conditions}

Let us consider a many body system consisting of $r$ Coulomb interacting 
species of charges. Its quantization is described by the transition amplitude

\be
{\cal Z}[\phi]=\frac{I[\phi]}{I[0]}=<\Omega_g, {\rm in} |\Omega_g, out>_\phi
\en

\noi where $|\Omega_g>$ is the ground state associated to the Hamiltonian in 
the absence of the external electrostatic field and it is assumed that the 
static field $\phi (\vec x)$ is adiabatically connected from $|\Omega_g, in>$
at the begining, further maintained during a long interval T and after
disconnected when the state becomes $|\Omega_g, out>$ [6].

\bi

The hamiltonian after introducing the field $\phi (\vec x ,t)$ is given by

\bq
H &=&\int d^3 x \sum^r_{s=1} \psi^+_s (\vec x) \hat O_s \psi_s 
(\vec x) -\int d^3x \Big( \sum^r_{s=1} q_s \psi^+_s (\vec x)\psi_s (\vec x)-
\rho_0 \Big) \phi (\vec x) + \nonumber \\
& &\frac{1}{2} N\left[ \int d^3x d^3y \sum^r_{s,t=1}q_s\psi^+_s (\vec x) \psi_s
(\vec x) \frac{(2-\delta_{st})}{\epsilon |\vec x - \vec y|}q_t\psi^+_t (\vec y)
\psi_t (\vec y) \right]
\eq

\noi where $N$ means normal ordering, $\hat Q_s$ are the kinetic one 
particle operators for each componet and $\rho_0$ a possibly needed jellium 
background density. The external field $\phi (\vec x)$ is assumed being 
generated by an electrostatic charge density through the Poisson
equation

\be
\phi (\vec x) = \int \frac{4\pi}{|\vec x - \vec y|} \rho (\vec y)
d^3 y .
\en

\bi

Let us consider a translationally invariant ground state of the system 
$|\Omega_g>$ and the adiabatic connection of a specific density 
$\rho (\vec x)$ under the above defined scheme. 
Then, the ground state to ground state 
transition amplitude given by ${\cal Z}$ in (1) should obey

\be
<\Omega_g,in|\Omega_g,out>=\exp
\Big(-\frac{i}{\hbar}\Delta E
[\phi ]T \Big) ,
\en

\noi with $\Delta E[\phi ]=E[\phi ]-E_g$ and $E[\phi ]$ and $E_g$ are the ground
state energies in presence and absence of the source $\rho $ respectively. 
For the generator of connected Green functions $W$ 
it follows at large L values 

\be
W [\phi (\vec x)]=-i\hbar \ln {\cal Z} [\phi (\vec x)] =- (E[\phi ]-E_g)T .
\en

Next, consider the search for the minimum of the mean value of the energy
in the state space under the restriction that the mean value of the
internally produced charge density takes a fixed static value $\rho_i 
(\vec x)$ that is

\be
\frac{<\Omega |\sum_{s} q_s \psi^+_s(\vec x)\psi_s(x)-\rho_0|\Omega >}{<\Omega
|\Omega >} = \rho_i (\vec x) .
\en

The usual normalization condition $<\Omega | \Omega >=1$ will be also
required.

\bi

Alternatively the extremum to be found can be obtained by using the method of 
Lagrange multipliers and minimizing the quantity

\be
<\Omega |H|\Omega >-\alpha <\Omega |\Omega >-\int d^3x \beta (\vec x)
<\Omega | \sum^r_{s=1} q_s \psi^+_s (\vec x) \psi_s (\vec x)-\rho_0 |\Omega >,
\en

\noi in which $\alpha$ and $\beta (\vec x)$ are to be determined from imposing
the normalization and mean field condition (6). The extremal of 
these expressions over arbitrary states leads to the following Schodinger 
equation for the solution $|\Omega >$.

\be
\Big( H -\int d^3x \beta (\vec x) \sum^r_{s=1} (q_s \psi^+_s (\vec x)
\psi_s (\vec x)-\rho_s) \Big) |\Omega >= \alpha |\Omega >.
\en

But, on another hand the state which is obtained after the adiabatic 
connection of a density $\rho (\vec x)$ is also abeying the
Schrodinger equation

\be
(H -\int d^3x \phi (\vec x) \Big(\sum_{s=1} q_s \psi^+_s (\vec x) \psi_s
(\vec x) -\rho_0 \Big) |\Omega_\phi >=E[\phi ] |\Omega_\phi > 
\en

Therefore, if the auxiliary charge $\rho (\vec x)$ is taken sufficiently 
small in order that no crossing of energy levels occurs in connecting it, the 
state $|\Omega_\phi >$ is also the ground state in presence of the source
$\rho (\vec x)$. Thus, minimal conditions under the constraints (6) and 
the normalization condition
are obeyed by the state $|\Omega_\phi >$ if the multiplier values are
selected to be

\bq
\alpha &=& E [\phi_{\rho_i} ] \nonumber \\
\beta (\vec x) &=& \phi_{\rho_i} (\vec x) 
\eq

\noi
where ~$\phi_{\rho_i}(\vec x)$ is such a field that forces the 
internal density \\
$$< \Omega | \sum_{s=1} q_s \psi^+_s (\vec x) \psi 
(\vec x)- \rho_0 |\Omega >$$
 
\noi
to be equal to $\rho_i (\vec x)$.  Here the central
point comes. Taking the scalar product of (9) with $|\Omega \phi_{\rho_i} >$ 
the minimal values for the mean energy, once the field $\phi(\vec x)$ is fixed,
takes the form

\bq
<H>_{\Omega_{\phi}} &=& <\Omega_{\phi} | H |\Omega_{\phi} > \nonumber \\
&=& \Delta E [\phi ]+\int d^3 x \rho_i (\vec x) \phi (\vec x) \nonumber \\
&=&-\frac{1}{T} (W[\phi ]-\int d^3x\rho_i(\vec x)\phi (\vec x))+E_g \nonumber \\
&=& E_g - \frac{1}{T} \Gamma [\rho_i ],
\eq

\noi where $\phi (\vec x)=\phi_{\rho_i}(\vec x)$ and the effective action 
$\Gamma [\rho_i]$ as a functional of any space time dependent field $\rho_i 
(x), x= (\vec x,t)$ is given by the Legendre transformation from the variables
$\phi (x)$ to the mean fields $\rho_i (x)$

\be
\Gamma [\rho_i ]=W[\phi (\vec x,t)]-\int dx \rho_i (\vec x,t) \phi (\vec x,t)
\en
\[ \rho_i (x) =\frac{\delta W}{\delta\phi (x)} , \]
\[ \int dz \frac{\delta\Gamma [\rho_i ]}{\delta \rho_i(x) \delta\rho_i (z)}|_{\rho_i=0} 
\frac{\delta W [\phi ]}{\delta\phi (z) \delta\phi (y)}|_{\phi=0}=- \delta^{(4)} (x-y) \] .

\bi

But, $<H>_{\Omega_{\phi}}$ in (11) is the minimum value of the mean energy under a 
restriction, therefore it should be necessarely greater than the ground state
energy which is the fully unrestricted extremal value. Thus, the stability 
condition for the assumed translational invariant ground state becomes

\be
<H>_{\Omega\phi} -E_g =-\frac{1}{T} \Gamma [\rho_i (\vec x)] \geq 0,
\en

\noi which requires the positivity of the effective potential under variation 
of the internal density $\rho_i (\vec x)$ 

\be
V[\rho_i(\vec x)]=-\lim_{T\to\infty}\left[ \frac{1}{T}\Gamma [\rho_i (x)]
\right] .
\en       

Thus, the functional formed by the second derivatives of $V$ should be a 
positive definite kernel. This is the stability condition obtained by Kirshnitz
[3] and it should be certainly obeyed. As it was discussed in [4] condition 
(14) also allows for negative values of the dielectric function. It will be
seen in what follows. Integrating the last of  relation (12) at vanishing
$\phi = \rho_i = 0$ over both time $t_x$ and $t_y$, dividing by the integration
interval $T$ and making use of the assumed translation invariance, it follows
for the second derivative of $V [\rho_i ]$

\bi

\[ \int d^3 z\Big(\int dt \Gamma^{(2)} (\vec x-\vec z,t)\Big) \int dt^\prime W^{(2)}
(\vec z-\vec y, t^\prime)=-\delta^{(3)} (\vec x-\vec y) , \]
\be
\int d^3z\frac{\delta^2 V [\rho_i ]}{\delta \rho_i (\vec x) \rho_i (\vec z)}|_{\rho_i=0} 
W^{(2)} (\vec z - \vec y, w=0)=\delta^{(3)} (\vec x - \vec y),
\en

\noi  in which

\be
\Gamma^{(2)} (x-y) = \frac{\delta^2 W}{\delta\rho_i (x) \delta\rho_i (y)}|_{\rho_i=0},
\en 
\be
W^{(2)} (x-y) = \frac{\delta^2 W}{\delta\phi (x) \delta\phi (y)}|_{\phi=0},
\en

\noi and $W^{(2)} (\vec x , \omega )$ is the temporal Fourier transform of 
$W^{(2)} (x)$.

\bi
y
That is, the second derivative kernel of the effective potential is the inverse
of the temporal Fourier transformation of $W^{(2)}$. The Fourier transform can
be evaluated as usual by introducing the completion formula of the state 
space between the two operators in the relevant time ordering. It produces

\bq
W^{(2)}(\vec x-\vec y,0)&=&\frac{1}{\hbar} \sum_{n\not = 0}2Re[<0|\sum^r_{s=1}
q_s \psi^+_s (\vec x) \psi_s  (\vec x) |n>  .\nonumber \\
& &\frac {1}{(\epsilon_n -\epsilon_0)} <n|\sum_{s=1} q_s 
\psi^+_s (\vec y) \psi_s (\vec y)|0>] ,
\eq

\noi
where $|0>=|\Omega_{\phi}>$  and $\epsilon_0=E_g $.

For the space Fourier transform of (18) ot follows

\be
W^{(2)} (\vec k,0)=\frac{1}{\hbar V} \sum_{n \not =0} 
\frac{2|<0|\hat\rho (\vec k)|n>|^2}{\epsilon_n -E_g}
\en

\noi with $\hat\rho_i (\vec k)$ equal to the total density operator $\sum_s 
q_s\psi^+_s(\vec x)\psi_s(\vec x)$. This expression for $W^{(2)}$ shows that the kernel 
$W^{(2)} (\vec x-\vec y)$ is positive definite whenever $E_g$ is really the 
ground state energy, that is $E_g \leq \epsilon_n$. Then, the kernel formed the by 
the second derivatives of the effective potential should be also positive definite 

\be
\int d^3x d^3y\rho_i (\vec x) \frac{\delta V [\rho_i ]}{\delta\rho_i 
(\vec x)\delta \rho_i (\vec y)} \Big|_{\rho_i =0} \rho_i (\vec y) \geq 0,
\en

\noi for an arbitrary density $\rho_i (\vec x)$.

\bi

Let us consider now further independent restrictions required by the stability 
of certain systems. For this purpose it will be made explicit below how the 
source $\rho_i (\vec x)$ can be considered as a fluctuation of the jellium 
charge density $\rho_0$. Note that the fixing of a definite jellium density 
is an external constraint on the system. In most of the real systems the 
charge density of the homogeneous jellium is fixed self consistenty by the 
interaction with the charged particles.

\bi

Consider the interaction of the system with a modified 
jellium in the form

\bq 
I=\frac{1}{2}\int d^3x d^3y ((\sum^r_{s=1} q_s \psi^+_s (\vec x) \psi_s
(\vec x) -\rho_0 -\rho (\vec x)) .\nonumber \\
\upsilon_\epsilon (\vec x-\vec y) (
\sum^r_{s=1} q_s \psi^+_s (\vec y) \psi_s (\vec y) -\rho_0 -\rho (\vec y)),
\eq     

\noi where the ordering of the density operators is not relevant if the 
Coulomb interaction is regularized through a parameter $\epsilon$
 in such a way that it vanish at
$\vec x - 
\vec y=0$. Also it can be assumed that $\upsilon_\epsilon$ remain being positive 
function $\upsilon_\epsilon (\vec x -\vec y) \geq 0$.

Relation (21) can be also written in the form

\bq
I&=&\int d^3x d^3y \sum^r_{s=1}q_s \psi^+_s (\vec x) \psi_s (\vec x)
\upsilon_\epsilon (\vec x - \vec y) \sum_{s=1} q_s \psi^+_s (\vec y) 
\psi_s (\vec y) \nonumber \\
&-& \sum^r_{s=1} q_s \psi^+_s (\vec x) \psi_s (\vec x) \upsilon_\epsilon 
(\vec x -\vec y) (\rho_0 + \rho (\vec y) )\nonumber \\
&+& \frac{1}{2} (\rho_0 + \rho) (\vec x) \upsilon_\epsilon (\vec x - \vec y)
(\rho + \rho (\vec x)) .
\eq 

But due to the constancy of $\rho_0$ it follows that on any state with fixed 
number of particles and in a thermodynamical  limit sense

\[ \int d^3x d^3y \sum^r_{s=1} \rho_s \psi^+_s (\vec x) \psi_s (\vec x)
\frac{1}{|x-y|} \rho_0 |\Omega>= \int d^3 \vec x \quad \hat Q \frac{1}{|\vec x|}
\rho_0 |\Omega> \]
\be
=\int d^3 x d^3 y \rho_0 \frac{1}{|x-y|} \rho_0 |\Omega>,
\en

\noi where $\hat Q$ is the total charge operator.

By also considering that the total jellium charge does not vary in the 
fluctuation, that is
\[ \int d^3 x \rho (\vec x)=0 \] ,
\noi it follows

\bq
I&=&\int d^3x d^3y \Big( \sum^r_{s=1} q_s \psi^+_s (\vec x) \psi (\vec x)
\frac{\upsilon_\epsilon (\vec x - \vec y)}{2} \sum^r_{s=1} q_s \psi^+_s 
(\vec y) \psi_s (\vec y) \\
&-& \Big(\sum^r_{s=1} q_s \psi^+_s (\vec x) \psi_s (\vec x)-\rho_0 \Big) 
\phi (\vec x)+\frac{1}{4\pi} \partial\phi (\vec x) \partial\phi 
(\vec x)+\frac{1}{2} \rho_0 \frac{1}{|\vec x - \vec y|} \rho_ 0 ).
\nonumber
\eq

Then, the Coulomb interaction operator of the system in (2) coincides with I 
apart from the necessary divergent self energy of the jellium plus the 
electrostatic self energy of the fluctuation. Clearly, the absence of the last 
term in (24) means that it is added in the energy $E [\phi ]$ in (9).

\bi

Finally let us consider the work done by the external systems in order to 
impose a constraint consisting in  the variation $\rho$ in the jellium density 
(or its equivalent: a fixation of evalue $\rho_i (\vec x)$ of the internal 
density).

\bi

Note that the work $\Sigma$ that should be performed to create, let say the 
mean  value of the internal density $\rho_i (\vec x)$
should be equal to the energy of the state after connecting 
$\phi_{\rho_i} (\vec x)$ plus the energy of the electrostatic field $\phi_{\rho_i} 
(\vec x)$ minus the ground state energy $E_g$. That is

\be
\Sigma [\phi_{\rho_i} ]= E [\phi_{\rho_i}] + \frac{1}{8\pi} \int d^3x
\partial \phi_{\rho_i} (\vec x) \partial \phi_{\rho_i} (\vec x) -E_g .
\en

\bi

Note, that this quantity differes radically from the effective potential
(14). From (5) it follows 

\bq
E^{(2)} [\phi_{\rho_i}] &=& \frac{\delta E (\phi )}{\delta \phi_{\rho_i} 
(\vec x) \delta \phi_{\rho_i} (\vec y) } \nonumber \\
&=& W^{(2)} (\vec x -\vec y , \omega = 0) \nonumber \\
=-\frac{1}{\hbar} \sum_{n\not =0}2& &\frac{Re[<0|\sum_s q_s\psi^+(\vec x)\psi 
(\vec x)|n><n|\sum_t q_t\psi^+_t(\vec y)\psi_t (\vec y)|0>]}
{\epsilon_n-\epsilon_0} .\nonumber \\
\eq  

Hence, the second derivatives of $\Sigma$ are given by

\begin{eqnarray}
\frac{\delta \Sigma[\phi_{\rho_i}]}{\delta \phi_{\rho_i} (\vec x) \delta\phi
(\vec y)}&=&-\frac{1}{4\pi} \nabla^2 \delta (\vec x -\vec y) - \frac{1}{\hbar}
\sum_{n\not =0} 2 {Re[<0|\sum_s q_s\psi^+_s (\vec x) \psi_s (\vec x)}|n> \nonumber \\
& & \frac {1}{\epsilon_n - 
\epsilon_0} <n|\sum_t q_t \psi^+_t (\vec y) \psi_t (\vec y) |0>] .
\end{eqnarray}

Further, its spatial Fourier transform takes the form 

\bq
\Sigma^{(2)} (\vec k) &=& \frac{ k^2}{4\pi} \Big( 1- 
\frac{\upsilon (\vec k)}{\hbar} \cdot \frac{2}{V} \sum_{n\not = 0} 
\frac{|<0|\hat\rho_{\vec k} |n>|^2}{\epsilon_n - \epsilon_0} \Big)  \nonumber \\
&=& \frac{\epsilon k^2}{4\pi} \cdot \frac{1}{\epsilon (\vec k)},
\eq

\noi
in which as, usual for the dielectric $\epsilon$, reponse R and polarization 
$P$ functions the following relations take place [7]

\bq
\frac{1}{\epsilon (\vec k)} &=& 1+ \upsilon (\vec k) R(\vec k) = 
\frac{1}{1-\upsilon (\vec k) P(\vec k)} , \nonumber \\
P (\vec k) &=& \frac{R (\vec k)}{1+\upsilon (k) R(k)} , \nonumber \\
R (\vec k) &=& \frac{P (\vec k)}{1-\upsilon (\vec k) P (\vec k)} \\
&=& \frac{2}{\hbar V} \cdot \sum_{n\not =0} \frac{|<0|\hat\rho_k|n>|^2}
{\epsilon_0 - \epsilon_n} . \nonumber
\eq

\bi

Notice that the stability condition following from the positivity of the
effective action is given by [3]

\be
-R(\vec k) =\frac{1}{\upsilon (\vec k)} \frac{1}{\epsilon (\vec k)} 
(\epsilon (\vec k)-1) \geq 0
\en

\noi which implies the inequalities

\be
\epsilon (\vec k)>1 \qquad\qquad {\rm or} \qquad\qquad \epsilon (\vec k)<0.
\en

Then, we arrived to the main point; the requirement that a positive work
is always needed to produce a definite value of the internal charge density 
$\rho_i (\vec k)$  leads to the positivity of of the dielectric function 
$\epsilon (\vec k)$ or its inverse. However, this condition should not be 
directly interpletable as a stability one.
In order to be so, the jellium should be a freely moving system which
homogeneity is self consistently determined by its strong Coulomb interaction 
with the particle components. Although, such should be the case in many 
systems, it is not the general one. For example, if the charges of the 
components completely cancels, then $\rho$ might describe only an 
external auxiliary charge. Then, the work to stablish it is not excluded that
could be negative. However, if the jellium is a freely moving and 
non-vanishing, it becames clear that the full system,that is particles + 
jellium, can diminish dinamically its energy by adopting an inhomogeneous 
configuration whenever $\epsilon (\vec k)$ can be negative.  

\bi

The existence of further  intrinsic 
physical conditions   requiring the positivity rule $\epsilon 
(\vec k)>0$ will be considered in the next section where the electromagnetic field
is considered as a dynamical one.

\textheight = 24truecm           
\textwidth = 16truecm 
\hoffset = -1.3truecm
\voffset = -2truecm

\section{$ \epsilon (\vec k)$ and dynamical photon fields}

In this section, an additional support for  the necessary positivity of the static
dielectric function  is given.  As it can be reminded, in last section such a 
condition  only followed  after assuming    that the jellium compensating charges 
were dynamical ones. Then, the stability of the whole system required  
$\epsilon(\vec k ) \geq 0$.  However, as it will be discussed below , it seems
that the internal stability of the charged system also requires that rule, no 
matter the character of the jellium dynamics. The argument follows from a 
treatment in which the electromagnetic field is considered as quantized.  Also the
 system will be described  at zero temperature by the euclidean quantum field
 theory limit  of the Matsubara  finite temperature theory [8]-[9].
 Let us consider the partition function of a charged system in interaction 
with the electromagnetic field as given by

\setcounter{equation}{32}
\begin{eqnarray}
{\cal Z} [j] &=& Tr\left[\exp (-\beta H)T\left(\exp\int -{1\over\hbar c}\int dx\ j_\mu (x)
\hat A_\mu (x)\right)\right] \nonumber \\ [3mm]
&=& {1\over N}\int{\cal D}\psi^+{\cal D}\psi{\cal D}A_\mu \exp\left[ S-{1\over\hbar c}\int
d^4x\ j_\mu (x) A_\mu (x)\right]
\end{eqnarray},

\noi
where the euclidean action for the charged system can be considered in the form  [9]

\begin{eqnarray}
S &=&{1\over\hbar c}\int^\beta_0 dx_4\int d^3x \left({\cal L}_f(\psi^+_s,\psi_s,A_\mu
)-{1\over 16\pi}\ F_{\mu\nu}F_{\mu\nu}\right.  \nonumber \\ 
&&\left. -{1\over 2\alpha}\ (\partial_\mu A_\mu
)^2\right) ,\quad s=1,\dots r ,
\end{eqnarray}

\noi in which
$F_{\mu \nu}=\partial_{\mu} A_\nu-\partial_\nu A_\mu$, $\psi^+_s$, $\psi_s$  
are the fermion  fields describing 
the $s=1,....r$ species of
 charged particles  and $ A_{\mu}$  is the electromagnetic field.
The operator $H$ is  the system hamiltonian  in the extended state space  including 
the non-physical photons as well as the auxiliary ghost particles.
The coordinate  $x_4$ is a real one and takes values in the interval  $(0,\beta)$ with 
$\beta=\hbar c/k T$.  The $\alpha$-dependent  term  in (34)
is associated  to the Fadeev-Popov quantization of the gauge electromagnetic field.
 In the case of the electromagnetic field, the contribution to the partition function
  (33) of the auxiliary ghost particles  compensating   the scalar and longitudinal
  unphysical degrees of freedom,  becomes a field independent
determinant. Here it  is assumed to be included in the normalization 
constant N in (33).
  For concreteness, the fermion part of the lagrangian can be taken to be 
the summ of r terms, each one corresponding  to the interaction of non-relativistic
 particles with the quantized electromagnetic field, that is  [10]

\begin{eqnarray}
{\cal L}_f &=&\sum^r_{s=1}\ \left( -\hbar c\psi^+_s\ {\partial\over\partial x_4}\
\psi_s-{1\over 2m_s}\psi^+_s \left(\vec p -{e\over c}\ \vec A(x)\right)^2\psi_s 
\right) \nonumber \\
& & - (\sum ^r_{s=1}q_s\psi^+_s(x)\psi_s(x)-\rho_0 )A_4(x) .
\end{eqnarray}

Expression (35) for ${\cal L}_f $ should be very accurate for the description of
 typical condensed matter  physical systems in which
the dynamics of the charged particles can be considered as non-relativistic.

The mean electromagnetic fields can be obtained through the logarithmic  
derivative of $\cal{Z}$ with respect to $j$ as

\begin{eqnarray}
\langle \hat A_\mu(x)\rangle &=& -\hbar c\ {\delta \ell n\ Z[j]\over \delta\ j_\mu (x)} =
{\delta\ W[j]\over \delta\ j_\mu (x)} \nonumber \\[3mm]
&=& {\int{\cal D}\psi^+{\cal D}\psi{\cal D}A_\mu\ A_\mu(x)
\exp [S-\int dx\ j_\mu A_\mu ]\over
\int{\cal D}\psi^+{\cal D}\psi{\cal D} A_\mu\exp [S-\int dx\ j_\mu A_\mu ]},
\end{eqnarray}

\noi
in which  the generating functional of the connected Green functions  $ W[j] $ 
    is defined by

\begin{equation}
W[j]=-\hbar c\ \ell n\ Z[j]
\end{equation}

Up to now, the finite temperature situation is yet being considered. Further, 
the $T=0$ limit will be assumed. 

It  can be remarked that the variety of subtle points  associated with the 
thermodynamical $T\neq 0$ situation  need for a more involved discussion  in 
order to properly  interpret the numerical or  experimental data and the existing 
viewpoints  [4]-[5]. These question will be considered elsewhere.

Now, it is possible to perform a Legendre transformation to consider 
the mean fields as the new independent variables in place of the sources
$j$, through

\begin{eqnarray}
\Gamma [A_\mu] &=& W[j]-\int dx\ j_\mu (x)A_\mu (x),\\
A_\mu (x) &=& {\delta\ W[j]\over \delta j_\mu (x)}.
\end{eqnarray}

Taking the derivative of (38) over $A$ it follows

\begin{equation}
{\delta\Gamma [A_\mu ]\over\delta\ A_\mu (x)} =-j_\mu (x)
\end{equation}

\noi
which after derivating over $j_{\nu}(y)$ and taking $j_{\mu}=A_{\mu}=0$
leads to

\begin{equation}
\int dz\ {\delta\Gamma [A]\over\delta\ A_\mu (x)\delta A_\alpha (z)}\Bigg\vert_{A=0}\
{\delta^2W[j]\over\delta j_\alpha (z)\delta j_\nu (y)}\Bigg\vert_{j=0} =
-\delta_{\mu\nu}\ \delta^{(4)}(x-y) .
\end{equation}

Up to now the sources and field $A_{\mu}$ have been considered as $x_4$ dependent 
in eucliden space. Then, let us consider the static limit  associated 
to equilibrium  configurations. In that case

\begin{equation}
\begin{array}{rl}
j_\mu &= (0,0,0,\rho (\vec x)) ,\\
A_\mu &= (0,0,0,\phi (\vec x)),
\end{array}
\end{equation}

\noi
where  it is assumed  that when the external source is purely electrostatic, the 
system is able only to generate  electrostatic mean fields. This 
is the case, in particular if parity invariance is not broken for 
example in the absence of external magnetic fields  
or other parity breaking effects.

The effective potential is then given by the limit

\begin{equation}
V[\varphi (\vec x)]=\lim_{L\to\infty}\left[{1\over L}\ \Gamma [0,0,0,\phi_L(\vec x,
x_1)]\right],
\end{equation}

\noi
in which  $\phi_L$ is taken as coinciding  with $\phi (\vec x )$ in $(0,L)$ and
diminishing to zero outside.

Now let us  consider as in the previous section the variational problem of finding 
the extremum of the mean energy within the subspace of physical states under 
a restriction for  a fixed mean field $A_{\mu}=(0,0,0, \phi(\vec{x} ))$ . 
Due to the extremality condition a Schrodinger equation should follow for the 
ground state in presence of a source  $ j_{\mu}=(0,0,0,\rho(\vec {x}))    $ 
in the form

\begin{equation}
\left( H +\int d^3x\ j_\mu (\vec x)\ \hat A_\mu (\vec x)
\right)\ \vert\Omega_j\rangle =
E[j]\vert\Omega_j\rangle ,
\end{equation}

\noi
where as before $ H $ is the system Hamiltonian in the 
extended space.

Taking the scalar product  with $|\Omega_j>$

\begin{equation}
\langle\Omega_j\vert H\vert\Omega_j\rangle -E_g=\left( E[j]-E_g-\int d^3x\ \rho (\vec
x)\varphi (\vec x)\right).
\end{equation}

But in the euclidean limit of the Matsubara theory  at 
great values of $ L $ it should follow

\begin{eqnarray}
-\hbar c\ \ell n\ {\cal Z}[j] &\cong&-
\hbar c\ \ell n\left[\exp\left( -{L\over\hbar c}\
E[j]\right)\right]\nonumber \\
W[0,0,0,\rho (\vec x)] &\cong& L\ E[j] .
\end{eqnarray}

 Then, in a similar way as in the previous section, and because $|\Omega_j>$ is
a conditional extremal within the physical states subspace,
  the following inequality should be valid

\begin{eqnarray}
\langle\Omega_j\vert H\vert\Omega_j\rangle -E_g
&=&\lim_{L\to\infty}\left[{1\over L}\left(
W[0,0,0,\rho (\vec x)-\int dx\ \rho (\vec x)\varphi_L (\vec x,x_4)\right)\right]
-E_g\nonumber
\\[3mm]
&=&\lim_{L\to\infty}\left({1\over L}\ \Gamma [0,0,0,\varphi_L (\vec x,x_4)]\right)
-E_g\nonumber \\[3mm]
&=& V[\varphi (\vec x )]-E_g\geq 0 .
\end{eqnarray}

Let us find now an expression for the kernel of the second derivatives of $ V $
which due to  (47) should be a positive definite one. The Fourier transform of the 
equation  (41 )  can be written as [8]

\begin{eqnarray}
\left( {k^2\over 4\pi}\ \delta_{\mu\alpha}\right. &-&\left.\left({1\over
4\pi}+{1\over\alpha}\right)\ k_\mu k_\alpha -\Pi_{\mu\alpha})k)\right)\ {\cal
D}_{\alpha\nu}(k) =-\delta_{\mu\nu},
\nonumber\\[3mm]
k &=& (\vec k,k_4),
\end{eqnarray}

\noi
where   $ \Pi_{\mu \nu}   $  is the polarization tensor  which due to gauge invariance  
satisfies the  tranversality condition

\begin{equation}
k_\mu\ \Pi_{\mu\nu}(k) =0
\end{equation}

 These relations allow  to evaluate  the Fourier tranform of the kernel  
$ \frac {\delta^2 V[\phi]} {\delta \phi({\vec x} )  \delta \phi({\vec y} )}$.
Note that the increment of a functional under a $x_4 $ independent 
increment  in the argument  is given by

\begin{equation}
\delta F[\phi ] =\int^\infty_{-\infty} dx_4\ {\delta\ F[\phi ]\over\delta\phi (\vec x,x_4)}\
\delta\phi (\vec x).
\end{equation}

Thus, after using translation invariance

\begin{eqnarray}
{\delta^2V[\phi ]\over \delta\phi (\vec x)\delta\phi (\vec y)}\Bigg\vert_{\phi =0} &=&
{1\over L}\int dx_4\ dy_4\ {\delta^2\Gamma [0,0,0,\phi ]\over
\delta\phi (\vec x,x_4)\ \delta\phi 
(\vec y,y_4)}\Bigg\vert_{\phi =0}\nonumber \\
[3mm]
&=& \int dx_4\ {\delta^2\Gamma [0,0,0,\phi ]\over
\delta\phi (\vec x-\vec y,x_4)\ \delta\phi (\vec 0,0)}
\Bigg\vert_{\phi=0}\nonumber\\[3mm]
&=&\Gamma^{(2)}(\vec x-\vec y,0),
\end{eqnarray}

\noi
where   $\Gamma^{(2)}(\vec{x}-\vec{y},k_4)   $   is the temporal Fourier 
transform  of the kernel

\begin{equation}
\Gamma^{(2)}(x-y)={\delta\Gamma [0,0,0,\phi ]\over
\delta\phi (x)\ \delta\phi (y)}\Bigg\vert_{\phi =0}.
\end{equation}

But, integrating relation  (41) over $y_4$  and taking 
the $ \mu=\nu=4 $ component leads to

\begin{equation}
\int d^3z\ \Gamma^{(2)}_{4\alpha}(\vec x-\vec z,0)\ {\cal D}_{\alpha 4}(\vec z-\vec
y,0)=-\delta^{(3)}(\vec x-\vec y),
\end{equation}

\noi
where as before  $ {\cal D}_{\mu \nu} $ is the 
temporal Fourier transform  of

\begin{equation}
{\delta^2W\over\delta j_\mu (x)\ \delta j_\nu (y)}\Bigg\vert_{j=0} .
\end{equation}

In momentum space it follows

\begin{equation}
\Gamma^{(2)}_{4\alpha} (\vec k,0)\ {\cal D}_{\alpha 4}(\vec k,0)=-1 .
\end{equation}

However, from the Dyson equation (48) in the static limit $k_4=0$ and
  $ \mu=\nu=4 $ also follows

\begin{equation}
\left({\vec k^2\over 4\pi}\ \delta_{4\alpha} -\Pi_{4\alpha}(\vec k,0)\right)\ {\cal
D}_{\alpha 4}(\vec k,0)=-1 .
\end{equation}

This relation can be further simplified  by noticing that by transversality 
and  the assumed 
parity invariance  $ \Pi_{\mu \nu} $ should have the  structure [8]

\begin{equation}
\Pi_{\mu\nu}(k)=\pi_t(k)\left(\delta_{\mu\nu}-{k_\mu k_\nu\over k^2}-\ell_\mu\ell_\nu
\right) +\pi_{44}(k)\ \ell_\mu\ell_\nu .
\end{equation}

\noi with

\begin{equation}
\ell_\mu ={1\over \sqrt{k^2}}\ \left(-{\vec k\over\vert\vec k\vert}\ k_4,\vert\vec k\vert
\right) ,\quad k^2=\vec k^2+k^2_4.
\end{equation}

Thus, in the static limit $k_4=0$

\begin{eqnarray}
\Pi_{4i} &=& \pi_t(k)\left(\delta_{4i}-{k_4k_i\over k^2}-\ell_4\ell_i\right)
+\pi_{44}(k)\ell_4\ell_i\nonumber\\
&=& 0,\qquad I=1,2,3 .
\end{eqnarray}

Therefore,  relation (56) reduces to

$$
\left({\vec k^2\over 4\pi}-\Pi_{44}(\vec k,0)\right)\ {\cal D}_{44}(\vec k,0)=-1,
$$
\noi
which gives  $ {\vec k}^2/{4 \pi} - \Pi_{44}(\vec{k},0)  $ as the 
Fourier transform  of the stability kernel $ \Gamma^{(2)}(\vec{v}-\vec{y},0) $ .
Henceforth, the stability condition takes the form

\begin{eqnarray*}
&&\int d^3x d^3y \,\phi (x)\ {\delta^2V[\phi ]\over\delta\phi (\vec x)\delta\phi (\vec
y)}\Bigg\vert_{\phi =0}\ \phi (\vec y)\  \nonumber\\
&&=\int d^3 k  \, \phi(\vec k)\ \left({\vec k^2\over 4\pi} -\Pi_{44}(\vec k,0)\right)\ \phi (k)\
 \nonumber\\
&&=\int d^3k\ \phi^*(k)\ {\vec k^2\over 4\pi}\ \left( 1-v(\vec k )P(\vec k)\right)\phi (k)
\nonumber\\
&&=\int d^3k\ \phi^*(k)\ {\vec k^2\over 4\pi}\in (\vec k)\phi (\vec k)\geq 0 ,
\end{eqnarray*}

\noi
for arbitrary fields  $ \phi(\vec{k}) $ and where  $\epsilon(\vec{k}) $ 
and $ P(\vec{k})$  are the dielectric 
 and polarization functions  and $ v(\vec{k})=4 \pi/ |\vec{k}|^2 $.  
Relation  (60) is the mentioned property.
That is,  the positivity of the dielectric function  as an additional 
stability condition  complementing the ones given in [3] and [4].

\bigskip

\section{Summary}

 It has been argued that the postivity of the dielectric function should be a 
necessary condition for the stability of a translational and parity  invariant 
ground state of a charged  many particle system  at zero temperature.  
The connection  of this condition  with the  alternative ones  in [3] and [4] 
was analyzed.   

In complementing the here presented viewpoint  it could be of interest 
to comment  on  the Wigner crystallization problem. It was  the main  
example  given in [3] and [4] for supporting the possibility for 
negative values of the dielectric function at zero
 temperature in a homogeneous state.  
The present discussion suggests that
in lowering the charge density  at $  T=0 $  the homogeneity 
is broken at a point of infinite screening  $ \epsilon -> \infty $ 
which is the unset 
for the appearing of negative values of $\epsilon (\vec k )  $. 
Then, if the positivity condition 
is correct, the translation invariance of the
 ground state is lost and the system pass  to an inhomogeneous state 
(local order, plasma wave, ...). In  such phases the notion of a dielectric 
constant  with only one wavevector argument $\vec{k}$ lost the meaning. Then, 
in further diminishing the density, the system could attain the Wigner 
crystallization phase. Therefore,  it seems that requiring as in [4],
that after lowering the density, the Wigner crystal can only appear as 
related  with a solution of a dispersion equation  $\epsilon(\vec k )=0  $ for a  
homogeneous state ( only one $\vec k$ argument of the dielectric function) 
is not necessarely valid.
But, this condition is the main argument given in [4] in order 
to associate the Wigner 
crystallization  problem as a physical system showing  negative values of 
$\epsilon(\vec k) $ at zero temperature.

Finally, the situation at finite temperature should be also commented. 
The present discussion and the previous work [5]  
indicate that a similar analysis 
could  be valid  at $ T\neq0$. This problem will be considered elsewhere.
In particular  in Ref.[12],  evaluations of the dielectric function  at $T\neq0$ 
for various classical plasma systems   have been presented.  The results show 
negative values  at inverse wavelength regions  of the order of the 
interparticle distances.  At such wavelengths  the notion of  a mean field 
as a statistically well defined thermodynamical internal parameter seems to loss sense.
Then, under the assumption that  a similar condition $\epsilon(\vec k)\geq 0 $ 
 could be found,
as implied by a minimal value of a thermodynamical potential, the 
negative values of the dielectric constant  obtained in classical plasma systems 
 could  be associated with anomalous non-gaussian probabilities 
 for fluctuations around the homogeneous field configuration [11]. 
It seems to me that such  a situation  could be interpreted perhaps, as a 
sort  of weak breaking  of the translational  invariance possibly associated to 
 some kind  of  measurable  spatial inhomogeneities. It should 
be noticed  that the thermal energies
related with 
the mentioned results  are substantially lower  than the 
coulomb energy at the mean interparticle distances. This observation leads  to 
the interest  in investigating  the local spatial order in 
 molecular dynamic calculations for such systems  
( the Monte Carlo configurations, not following the equation of motion, may 
introduce an artificial homogeneization in the results).  A further investigation 
of this more involved  $ T\neq 0  $ case will be considered elesewhere.  

\bi

\noi {\large \bf Acknowledgments}

 I would like to thank the helpful remarks and comments of Profs.
N.H. March and G. Mussardo.
 In addition I should  express my gratitude to the Associate Programme
of the Abdus Salam International Centre for Theoretical Physics and
CONACyT of Mexico for the support in the realization of this work.

\newpage

\noi {\large \bf References}

\begin{itemize}
\item[1.-] M.L. Cohen and P.W. Anderson (1972). "Comments in the Maximum
	   Superconducting Transition Temperature", in Superconductivity
	   in d- and f- Band Metals, edited by D.H. Douglass (AIP, New York)
	   p. 17.
\item[2.-] P.C. Martin, Phys. Rev. 161, (1967) 143. 
\item[3.-] D.A. Kirshnitz, Sov. Phys. Usp. 19 (1976) 530.
\item[4.-] O.V. Dolgov, D.A. Kirshnitz and F.G. Maksimov, Rev. Mod.
	   Phys. 53 (1981) 81. 
\item[5.-] A. Cabo and A. P\'erez-Mart\ii nez, Phys. Lett. A 222 (1996) 263. 
\item[6.-] S. Weinberg, "The Quantum Theory of Fields", Vol. II; (Cambridge
	   Univ. Press, 1996) p. 72. 
\item[7.-] D. Pines, "The Many Body Problem", (W.A. Benjamin, Inc., 1973). 
\item[8.-] E. S. Fradkin, "Quantum Field Theory and Hydrodynamics", Ed. by 
                D.V. Skobel'tsyn, Proceedings of the P.N. Lebedev Physical Institute,
                 Vol. 29 (Consultant Bureau, 1967).
\item[9.-] C. W. Bernard, Phys. Rev.  D9 (1974)3312. 
\item[10.-] W.P. Healey, 'Non-Relativistic Quantum Electrodynamics", (Academic 
               Press, 1982).
\item[11.-] L.D. Landau and E.M. Liftshitz, 'Statistical Physics", Part I (Pergamon Press,
                 1980). 	
\item[12.-] A. Fasolino, A. M. Parrinelo and M. Tosi, Phys. Lett. A66 (1978)119.      

\end{itemize} 
\end{document}